\def\be{\begin{equation}}
\def\te{\end{equation}}
\def\bea{\begin{eqnarray}}
\def\nn{\nonumber}
\def\tea{\end{eqnarray}}

\def\d{\delta}

\def\h{\eta}

\def\l{\lambda}

\def\q{\theta}

\newskip\humongous \humongous=0pt plus 1000pt minus 1000pt

\newif\ifdtup

\def\nli{\nu_L^i}
\def\nri{\nu_R^i}

\newcommand{\Ubar}{\overline U}
\newcommand{\Dbar}{\overline D}
\newcommand{\Ebar}{\overline E}

\documentstyle[12pt]{article}


\begin{document}
\begin{titlepage}
\begin{flushright}
{\large
UMD-PP/97-5 \\
hep-ph/9607446\\
July 1996\\
}
\end{flushright}
\vskip 1cm
\begin{center}
{\Large\bf The essential role of string$-$derived symmetries in} \\ \vskip .2cm
{\Large\bf ensuring proton$-$stability and light
neutrino masses\footnote{Research supported by NSF grant No.
PHY-9119745. Email: {\tt pati@umdhep.umd.edu}}}
\vskip 1cm
{\large
Jogesh C. Pati\\}
\vskip 0.5cm
{\large\sl Department of Physics\\
University of Maryland\\
College Park, MD~20742\\}
\end{center}
\vskip .5cm
\begin{abstract}
The paper addresses the problem of suppressing naturally
the unsafe $d=4$ as well as the color-triplet mediated and/or
gravity-linked $d=5$ proton-decay operators, which
generically arise in SUSY-unification. It also attempts to give
light masses to the
neutrinos, of the type suggested by current experiments. It is noted
that neither the symmetries in $SO(10)$, nor those in $E_6$,
suffice for the purpose -- especially in the matter of
suppressing naturally the $d=5$ proton-decay operators. By contrast,
it is shown that a certain {\it string-derived symmetry}, which
cannot arise within conventional grand unification, but which
does arise within a class of three-generation string-solutions,
suffices, in conjuction with $B-L$, to
safeguard proton-stability
from all potential dangers, including those which may arise
through higher dimensional operators and the color-triplets in the
infinite tower of states. At the same time, the symmetry in question
permits neutrinos to acquire appropriate masses. This
shows that {\it string theory plays an esential role in
ensuring natural consistency of SUSY-unification
with two low-energy observations -- proton-stability and light
masses for the neutrinos}. The correlation between the masses of
the extra $Z'$-boson (or bosons), which arise in these models, and
proton-decay rate is noted.

\end{abstract}
\end{titlepage}

\section{Introduction}
\baselineskip=20pt
While supersymmetry is an essential ingredient for higher unification, it
is known that it poses the generic problem of rapid proton decay \cite{1}.
This is because, in accord with the standard model gauge symmetry
$SU(2)_L \times U(1)_Y \times SU(3)^C$, a supersymmetric theory in general
permits, in contrast to non-supersymmetric ones, dimension $4$ and dimension
$5$ operators which violate baryon and lepton numbers.
Using standard notations, the operators in question which may arise in the
superpotential are as follows:
\bea
W &=& [\eta_1 \Ubar \, \Dbar \, \Dbar  + \eta_2 Q L \Dbar + \eta_3 L L
\Ebar ] \nn\\
&+& [\l _1 QQQL + \l _2 \Ubar \, \Ubar \, \Dbar \, \Ebar  +
\l _3 LLH_2 H_2]/M.
\tea
Here, generation, $SU(2)_L$ and $SU(3)^C$ indices are suppressed.
$M$ denotes a characteristic mass scale.
The first two terms of $d=4$, jointly, as well as the $d=5$ terms of
strengths
$\lambda_1$ and $\lambda_2$, individually,
induce $\Delta(B-L) = 0$ proton decay with amplitudes
$\sim \h _1 \h _2/m_{\tilde
{q}}^{2}$ and $(\l _{1,2}/M)(\d )$ respectively, where $\d $
represents a loop-factor. Experimental limits on proton
lifetime turns out to impose the constraints:
$\h _1 \h _2 \leq 10^{-24}$ and $(\l _{1,2}/M)\leq 10^{-25}$ GeV$^{-1}$
\cite{3}.
Thus, even if $M \sim M_{string} \sim 10^{18}$ GeV, we must
have  $\l _{1,2} \leq 10^{-7}$,
so that proton lifetime will be in accord with experimental limits.

Renormalizable, supersymmetric standard-like and $SU(5)$ \cite{4}
models can be constructed
so as to avoid,
{\it by choice}, the $d=4$ operators (i.e. the $\h _{1,2,3}$-terms) by
imposing a discrete or a multiplicative
$R$-parity symmetry: $R\equiv (-1)^{3(B-L)}$, or more
naturally, by gauging $B-L$, as in ${\cal G}_{224}
\equiv SU(2)_L\times SU(2)_R\times SU(4)^C$ \cite{5} or $SO(10)$ \cite{6}.
Such resolutions,
however, do not in general suffice if we
permit higher dimensional operators and
intermediate scale VEVs of fields which violate $(B-L)$
and $R$-parity (see below). Besides, $B-L$ can not
provide any protection against the $d=5$ operators given
by the $\l_1$ and $\l_2$ - terms, which conserve $B-L$. These
operators are, however, expected to be present in any
theory linked with gravity, e.g. a superstring theory,
unless they are forbidden by some new symmetry.

For SUSY grand unification models,
there is the additional problem that the exchange of
color-triplet Higgsinos which
occur as partners of electroweak doublets (as in ${\bf 5}
+{\bf \overline{5}}$
of $SU(5)$) induce $d=5$ proton-decay operators \cite{1}.
Thus, allowing for suppression of $\lambda_1$ and
$\lambda_2$ (by about $10^{-8}$) due to the smallness
of the Yukawa couplings, the color-triplets still
need to be
superheavy ($\geq 10^{17}$ GeV) to ensure proton-stability
\cite{3},
while their doublet partners must be light ($\leq 1$ TeV).
This is the {\it generic problem of doublet-triplet splitting}
that faces all SUSY GUTS. Solutions to this problem needing
either unnatural fine-tuning
as in SUSY $SU(5)$ \cite{4},
or suitable {\it choice} of large number and/or large size
Higgs multiplets and discrete
symmetries
as in SUSY $SO(10)$ \cite{7} and missing-partner
models \cite{8}, are technically feasible. They, however, do not
seem to be
compelling because they  have been invented
for the sole purpose of suppressing proton-decay, without a
deeper reason.
Furthermore, such solutions are
not easy to realize, and to date have not been realized, in
{\it string-derived} grand unified theories \cite{9}.

These considerations show that, in the
context of supersymmetry, the extraordinary stability of the
proton is in fact surprising. As such it deserves a natural explanation. Rather
than being merely accomodated, it ought to emerge as a 
{\it compelling feature}, owing to symmetries of the underlying
theory, which should forbid, or adequately supress, the unsafe operators
in Eq. (1). As discussed below, the task of finding such symmetries
becomes even harder, if one wishes to assign
 non-vanishing light masses ($\leq$ few eV) to
neutrinos. The purpose of this letter is to propose a class of solutions,
within supersymmetric theories, which {\it (a)} naturally 
ensure proton-stability, to the extent desired, and
{\it (b)} simultaneously permit neutrinos to acquire light masses, 
of a nature that
is relevant to current experiments \cite{10}. These solutions need 
\underline{either}
$I_{3R}$ and $B-L$ as {\it separate} gauge
symmetries, as well as {\it one extra
abelian symmetry} that lies beyond even $E_6$ \cite{6};
\underline{or} the weak hypercharge
$Y$ ($=I_{3R} + (B-L)/2$) accompanied by {\it two extra symmetries}
beyond those of $E_6$. The interesting point is that while
 the extra symmetries in question can not arise within conventional
grand unification models, including $E_6$, they do
arise within a class of
string-derived three generation solutions. This in turn
provides {\it a strong motivation}  for symmetries
of string-origin. The extra symmetries
lead to extra $Z'$-bosons, whose currents would bear
the hallmark of string theories. It turns out that there is an interesting
correlation between the masses of the $Z'$-bosons and observability
of proton decay.

\section{The need for symmetries beyond $SO(10)$ and $E_6$}
\baselineskip=20pt
In what follows, we assume that operators (with d $\geq 4$), scaled
by Planck or string scale-mass, that respect all symmetries, exist
in the effective superpotential of
any theory which is linked to gravity, like a superstring
theory \cite{11,12}. For reasons discussed before, the class of
theories -- string-derived or not --
which contains $B-L$ as in  
${\cal G}_{2311} \equiv 
SU(2)_L\times SU(3)^C\times U(1)_{I_{3R}}\times U(1)_{B-L}$ as a
symmetry, the $d=4$ operators in Eq.($1$) are naturally forbidden.
They can {\it in general} appear however
through non-renormalizable operators if there exist VEVs of
fields which violate
$B-L$. This is where neutrino-masses
become relevant. The familiar see-saw mechanism
\cite{13} that provides the simplest reason for known neutrinos
$\nli $'s to be so light assigns {\it heavy Majorana masses}
$M_R^i$
 to the right-handed neutrinos $\nri $, and thereby light
 masses
$m_L^i\sim
(m_D^i)^2/M_R^i$ to the left-handed ones, where
$m_D^i$ denotes a typical Dirac mass for the $i$th
neutrino. These masses would have just the right pattern to be
relevant to the neutrino-oscillation experiments \cite{10,14}
and to $\nu_{\tau}$
being
hot dark matter, with $m_L^i\sim (10^{-8}, 3\times 10^{-3}$
and $1$-$10$) eV for $i=e,\mu,\tau$, {\it if} $M_R^i\sim 10^{12}$ GeV
(within a factor of $10$).
Generating heavy Majorana masses for $\nu_R$'s, however, needs
spontaneous violation of $B-L$ at a heavy intermediate scale.

If $B-L$ is violated by the VEV of a field by two units, an
effective $R$-parity would
still survive \cite{15}, which would forbid the $d=4$ operators. That
is precisely the case for the multiplet
126 of $SO(10)$ or $(1,3,\overline{10})$ of ${\cal G}_{224}$,
which have commonly been used \cite{13} to give Majorana masses to
$\nu_R$'s. Recent works show, however, that $126$ and very likely
$(1,3,\overline{10})$, as well,  are hard
-- perhaps impossible -- to obtain in string 
theories \cite{16}. We, therefore, assume that this constraint holds.
It will become clear,
however, that as long as we demand safety from both $d=4$ {\it and}
$d=5$ operators, our conclusion as regards
the need for symmetries beyond $E_6$, would hold even if we give up
this assumption.

Without 126 of Higgs, $\nu_R$'s
can still acquire heavy Majorana
masses utilizing product of VEVs of 
$s$neutrino-like fields $\widetilde{\overline{N_R}}$ and
$\widetilde{N'_L}$, which belong to 
$16_H$ and $\overline{16}_H$ respectively. (as in Ref. \cite{17}, 
see also \cite{18}.)
In this case, an effective operator of the form $ 16 \cdot 16 \cdot
\overline{16}_H.
\overline{16}_H/M$ in $W$, that is allowed by $SO(10)$,
would induce a Majorana mass  $(\overline{\nu}_R C^{-1}
\overline{\nu}_R^T)
(\langle\widetilde{N_L'}\rangle\langle\widetilde{N_L'}\rangle/M) 
+ hc\,$ of magnitude
$M_R\sim 10^{12.5}$ GeV, as desired, for
$\langle\widetilde{N'_L}\rangle \sim 10^{15.5}$ GeV and $M\sim 10^{18}$ GeV.
However, consistent with $SO(10)$ symmetry
and therefore its subgroups, one
can have
an effective $d=5$ operator in the superpotential
$ 16^a \cdot 16^b \cdot 16^c \cdot 16_H/M $.
%where $M\sim M_{st}\sim 10^{18}$ GeV (say). 
This would
induce the terms $\Ubar _R \Dbar _R
\Dbar _R \langle\widetilde{\overline{N_R}}\rangle/M$ and  $QL\Dbar
\langle\widetilde{\overline{N_R}}\rangle/M$
in $W$ (see Eq.($1$)) with strengths $\sim \langle\widetilde{\overline{N_R}}\rangle/M
\sim 10^{15.5}/10^{18} \sim 10^{-2.5}$, which would lead to unacceptably short
proton lifetime $\sim10^{-6}$ yrs. \cite{19}. 
We thus see that, 
without having the $126$ or $(1, 3, \overline{10})$
 of Higgs,
{\it $B-L$ and therefore $SO(10)$ does not suffice
 to suppress even the $d=4$ - operators adequately
 while giving appropriate masses to neutrinos}. 
As mentioned before, $B-L$ does not of course prevent
the $d=5$, $\l_1$
and $\l_2$ - terms, regardless of the Higgs spectrum, because
these terms conserve $B-L$.

To cure the situation mentioned above, we need to utilize
symmetries beyond those of $SO(10)$. Consider first
the presence of at least one extra $U(1)$ beyond $SO(10)$ of the type
available in $E_6$, i.e. $E_6\rightarrow SO(10)\times U(1)_{\psi}$,
 under which
$27$ of $E_6$ branches into $(16_1 + 10_{-2} + 1_4)$, where $16$ contains
$(Q, L \mid \overline{U_R}, \overline{D_R}, 
\overline{E_R}, \overline{\nu_R})$, with $Q_{\psi}
= +1$; while $10$
contains the two Higgs doublets
$(H_1, H_2)^{(0, -2)}$ and a color-triplet
and an anti-triplet
$(H^{(-2/3, -2)}_{\bf 3} + H^{'(2/3, -2)}_{3^{\ast}})$, where the
superscripts denote
$(B-L, Q_{\psi})$. Assume that the symmetry in the observable sector just
below the Planck scale is of the form:
\be
{\cal G}_{st} = [{\cal G}_{fc} \subseteq SO(10)]
\times \hat{U}(1)_{\psi} \times [U(1)'s].
\te
It is instructive to first assume
that $\hat{U}(1)_{\psi} = U(1)_{\psi}$ of $E_6$ \cite{20}
and ignore all the other $U(1)$'s.
Ignoring the doublet-triplet splitting problem for a moment, 
we allow the 
flavor-color symmetry ${\cal G}_{fc}$ to be as big as $SO(10)$.
The properties of the operators in $W$ given in Eq.($1$), and
of the fields
$\widetilde{\overline{N_R}}$, $(H_1, H_2)$ and the singlet $\chi\subset 27$,
under the charges $Y$, $I_{3R}$, $B-L$, $Q_{\psi}$ and
$Q_{T}\equiv Q_{\psi} - (B-L)$, are shown in Table $1$.
\begin{table}
\centering
\begin{tabular}{|c|c|c|c|c|c|}
\hline
Operators & $I_{3R}$ & $B-L$ & $Y$ & $Q_{\psi}$ & $Q_{T}$ \\
\hline
 & & & & & \\
$\Ubar \,\Dbar \,\Dbar$, $QL\Dbar$ & $1/2$ & -$1$ & $0$ & 3 & $4$ \\
$LL\overline{E}$ & $1/2$ & -$1$ & $0$ & $3$ & $4$ \\
$QQQL/M$ & $0$ & $0$ & $0$ & $4$ & $4$ \\
$\overline{U}\,\overline{U}\,\overline{D}\,\overline{E}/M$
& $0$ & $0$ & $0$ & $4$ & $4$ \\
$LLH_2H_2/M$ & $1$ & -$2$ & $0$ & -$2$ & $0$  \\
 & & & & & \\ \hline
 & & & & & \\
$\widetilde{\overline{N_R}}$ & -$1/2$ & $1$ & $0$ & $1$ & $0$ \\
$(H_1, H_2)$ & (-$1/2, 1/2)$ & $0$ & (-$1/2, 1/2)$ & -$2$ & -$2$ \\
$\chi$ & $0$ & $0$ & $0$ & $4$ & $4$ \\
 & & & & & \\ \hline
\end{tabular}
\caption{}
\end{table}
We see that the $d=4$ operators ($\eta_i$ -terms) are forbidden by
$B-L$, as well as by $Q_{\psi}$ and $Q_T$. Furthermore, note that when
$\widetilde{\overline{N_R}}\subset 16$ and
$\widetilde{N'_L}\subset\overline{16}$ acquire VEV, and give
Majorana masses to the $\nu_R$'s, the charges $I_{3R}$,
$B-L$ as well as $Q_{\psi}$ are broken, {\it but $Y$ and $Q_T$ are preserved}.
Now $Q_T$ would be violated by the VEVs of $(H_1, H_2) \sim
200$ GeV and of the singlets $\chi^{(27)}$ and
$\overline{\chi}^{(\overline{27})}$. 
Assume that $\chi$ and $\overline{\chi}$
acquire VEVs $\sim 1$ TeV through a radiative mechanism,
utilizing Yukawa interactions, analogous to
$(H_1, H_2)$. The $d=4$ operators can be induced
through nonrenormalizable terms of the type 
$16 \cdot 16 \cdot 16 \cdot [\langle\widetilde{
\overline{N_R}}\subset 16\rangle/M]. 
[\langle10\rangle\langle10\rangle/M^2$ or $\langle\overline{\chi}\subset
\overline{27}\rangle/M]$, where the effective couplings 
respect $SO(10)$ and $U(1)_{\psi}$.
Thus we get $\eta_i \leq (10^{15.5}/
10^{18})(1$ TeV$/10^{18}$ GeV)$\sim 10^{-18}$, which is below the
limit of $\eta_1 \eta_2 \leq 10^{-24}$. Thus, $B-L$
{\it and} $Q_{\psi}$, arising within $E_6$, suffice to control the
$d=4$ operators adequately, while permitting neutrinos to have
desired masses. 

Next consider the
$LLH_2H_2$-term. While it violates $I_{3R}$, $B-L$ and $Q_{\psi}$, it is
the only term that is allowed by $Q_T$. Such a term can arise through an 
effective interaction of the form $16 \cdot 16 \cdot (H_2\subset10)^2
\cdot \langle\widetilde{\overline{N_R}}\subset 16\rangle^2/M^3$, and 
thus with a strength
$\sim 10^{-5} \cdot (10^{18}$GeV)$^{-1}$, which is far below the limits
obtained from $\nu$-less double
$\beta$-decay.

Although the two $d=5$ 
operators $QQQL/M$ and 
$\overline{U}\,\overline{U}\,\overline{D}\,\overline{E}/M$
are forbidden by $Q_{\psi}$ and $Q_T$, the problem of
these two operators still arises as follows. Even for a
broken $E_6$-theory, possessing $U(1)_{\psi}$-symmetry, the
color-triplets $H_3$ and $H'_{3^\ast}$ of 27 still exist in the
spectrum. They are in fact needed to cancel
the anomalies in $U(1)_{\psi}^3$ and $SU(3)^2 \times U(1)_{\psi}$
etc. They acquire masses of the form $M_3 H_3 H'_{3^\ast}
+ hc$ through the VEV of singlet $\langle \chi \rangle$ which
breaks $Q_{\psi}$ and $Q_T$ by
four units. With such a mass term, the exchange of these
triplets would induce $d=5$ proton-decay operators, just as it does
for SUSY $SU(5)$ and $SO(10)$. We are then back to facing either
the problem of doublet-triplet splitting (i.e. why $M_3 \geq 10^{17}$ GeV)
or that of rapid proton-decay (for $M_3 \sim 1$ TeV). In this sense,
while the
$E_6$-framework, with $U(1)_{\psi}$, can adequately control the
$d=4$ operators and give appropriate masses to the neutrinos
(which $SO(10)$ cannot), it does not suffice to
control the $d=5$ operators, owing to the presence of
color-triplets. As we discuss below, this is where string-derived
solutions help in preserving the benefits of a $Q_{\psi}$-like charge,
while naturally eliminating the dangerous color-triplets.

\noindent {\bf Doublet-Triplet Splitting In String Theories: A
Preference For
Standard-like Symmetries over GUTS:}
While the problem of
doublet-triplet splitting does not have a compelling solution
within SUSY GUTS and has not been resolved within string-derived
GUTS \cite{9},  it can be solved quite simply
within string-derived
standard-like \cite{21,22} or the ${\cal G}_{224}$-models \cite{17},
because in these models, the electroweak doublets 
are naturally decoupled from the
color-triplets after string-compactification. As a result,
invariably, 
the same set of boundary conditions
(analogous to ``Wilson lines'')
which break $SO(10)$ into a standard-like
gauge symmetry such as ${\cal G}_{2311}$, either project out,
by GSO projections, all
 color-triplets $H_3$ and $H_{3^\ast}^{'}$ from
the
``massless''- spectrum \cite{22}, or yield some color-triplets
with extra $U(1)$ - charges which make them harmless
\cite{21}, because they can
not have Yukawa couplings with quarks and leptons.
In these models, the doublet triplet splitting problem is
thus solved from the start, because the {\it dangerous} color -
triplets simply do not appear in the massless spectrum \cite{24}.

At the same time, owing to constraints of string theories,
the coupling unification relations hold \cite{23} for the
standard-like or ${\cal G}_{224}$-models, just as in GUT.
Furthermore, close to realistic models have been derived
from string theories only in the context of such standard-like
\cite{21,22}, flipped $SU(5) \times U(1)$ \cite{24} 
and ${\cal G}_{224}$ models \cite{17},
but not yet for GUTS.
For these reasons,
we will consider string-derived non-GUT models, 
as opposed to GUT-models,
{\it as the  prototype
of a future realistic string model}, and use them as a {\it guide}
to ensure {\it (a)} proton -
stability and {\it (b)} light neutrino masses.

Now, if we wish to preserve the benefits of the charge $Q_{\psi}$
(noted before), and still eliminate the color-triplets
as mentioned above, there would appear to be a problem, because,
without the color-triplets,
the incomplete subset consisting of 
$\{16_1 + (2,2,1)_{-2} + 1_4\}\subset 27$ of $E_6$
would lead to anomalies in $U(1)_{\psi}^3$,
$SU(3)^2\times U(1)_{\psi}$ etc. This is where symmetries of
string-origin come to the rescue.

\section{The crucial role of string-derived symmetries}
\baselineskip=20pt
The problem of anomalies (noted above) is cured within string
theories in a variety of ways. For instance,
new states beyond those in the $E_6$-spectrum invariably appear
in the string-massless sector which
contribute toward the cancellation of anomalies, and only certain
{\it combinations} of generators become
{\it anomaly-free}. We must then examine whether such
anomaly-free combinations
can help achieve our goals.
To proceed further, we need to focus on some specific
solutions. For this purpose, we choose to explore here
the class of string-derived three generation models, obtained
in Refs.\cite{21} and \cite{22}, which is as close
to being realistic as any other such model that exists in the
literature (see e.g. Refs. \cite{17} and \cite{24}). In particular, they seem capable of generating
qualitatively the right texture for fermion mass-matrices and CKM mixings.
We stress, however, that the essential feature of our
solution, relying primarily on the existence of extra
symmetries analogous to $U(1)_{\psi}$,
is likely to emerge
in a much larger class of string-derived solutions.

We refer the reader to Refs. \cite{21} and \cite{22} and references
therein for the procedure of choosing string-boundary
conditions, applying GSO projections, and deriving the
effective low-energy theory.
After the application of all GSO projections, the gauge symmetry
of the models developed in these references, at the string
scale, is given by:
\be
{\cal G}_{st} = [SU(2)_L \times SU(3)^C \times U(1)_{I_{3R}}
\times U(1)_{B-L}]
\times [G_M = \prod_{i=1}^{6} U(1)_{i}] \times G_H.
\te
Here, $U(1)_i$ denote six horizontal-symmetry charges which
act non-trivially on the three families and distinguish between
them.  In the models
of Refs. \cite{21}, \cite{22}, $G_H = SU(5)_H \times SU(3)_H \times
U(1)_H^2$. There exists ``hidden'' matter which couples
to $G_H$ and also $U(1)_i$. Thus the
gauge interactions
of the sector $G_M = [U(1)]^6$ serve as the {\it messenger} between
the hidden and the observable matter.
The form of $G_M$ varies from model to model, but its
occurence seems to be a {\it generic} feature
(see e.g. Refs. \cite{24}, \cite{17}, \cite{18}).

A partial list of the massless states for the solution
derived in Ref. \cite{21}, together with the associated $U(1)_i$-charges,
is given in Table 2.  We have not exhibited
a host of other states including {\it (a)} 10 pairs of $SO(10)$-singlets,
with $U(1)_i$-charges, {\it (b)} three universal
singlets $\xi_{1,2,3}$, and
{\it (c)} states with fractional charges which either get
superheavy or get confined \cite{25}. The table reveals the
following features:

\noindent {\it (i)} There are three families
of quarks and leptons (1, 2 and 3), each with $16$ components,
including $\overline{\nu_R}$. Their quantum numbers under the symmetries
belonging to $SO(10)$
are standard and are thus not shown. Note that the
$U(1)_i$ charges differ from one family to the other. There are
also three families of hidden sector multiplets $V_i$, $\overline{V}_i$,
$T_i$ and $\overline{T}_i$ which possess $U(1)_i$-charges.

\noindent {\it (ii)}
The charge $Q_1$ has the same value ($\frac{1}{2}$) for all
sixteen members of family $1$, similarly $Q_2$ and $Q_3$ for
families 2 and 3 respectively. In fact, barring a normalization
difference of a factor of 2, {\it the sum $Q_+ \equiv Q_1 + Q_2 +
Q_3$ acts on the three families and on the three Higgs doublets
$\overline{h_1}$, $\overline{h_2}$ and $\overline{h_3}$ in the
same way as the $Q_{\psi}$ of $E_6$ introduced before}. The
analogy, however, stops there, because the solution has
additional Higgs doublets (see table) and also because
there is only one pair of color triplets ($D_{45}, \overline{D}_{45}$)
instead of three. Furthermore, the pair
($D_{45}, \overline{D}_{45}$) is {\it vector-like} with
opposite $Q_+$-charges, while
($H_3, H'_{3^\ast}$), belonging
to $27$ of $E_6$, have the same $Q_{\psi}$-charge.
In fact the pair ($D_{45}, \overline{D}_{45}$) can have
an invariant mass conserving all $Q_i$-charges, but ($H_3, H'_{3^\ast}$)
can not.

\noindent {\it (iii)} It is easy to see that owing to different
$U(1)_i$-charges, the color-triplets
$D_{45}$ and  $\overline{D}_{45}$ (in contrast to $H_3$ and
 $H'_{3^\ast}$) can not have allowed Yukawa couplings to
 ($qq$) and ($ql$) - pairs. Thus, as mentioned before, they
 can not mediate proton decay.

\noindent {\it (iv)} Note that the solution yields altogether
four pairs of electroweak Higgs doublets: ($h_1, h_2, h_3,
h_{45}$) and ($\overline{h_1}, \overline{h_2}, \overline{h_3},
\overline{h_{45}}$). It has been shown \cite{21} that only one
pair -- i.e. $\overline{h_1}$ or $\overline{h_2}$ and
$h_{45}$ -- remains light, while the others acquire superheavy
or intermediate scale masses. The cubic level
superpotential has the form
\cite{21}:
\bea
 W &=& \left\{ \sum_{i=1}^3 \Ubar _i Q_i \overline{h_i}
+ \sum_{i=1}^3 N_{Li}^C L_i \overline{h_i} + \sum_{i<j} h_i
\overline{h_j} \overline{\phi_{ij}} + \sum_{i<j}
\overline{h_i} h_j \phi_{ij}\right\} \nn \\
& &+ {\cal O}(\phi^3) + \left\{ \frac{\xi_3}{2}
(D_{45}\overline{D}_{45} + h_{45}\overline{h_{45}} +
{\cal O}(\phi^2) \right\}+ \cdots,
\tea
 where $\phi$'s denote $SO(10)$ - singlets,
possessing $U(1)_i$-charges. Note that, owing to differing
$U(1)_i$-charges, the three families have Yukawa couplings
with three distinct Higgs doublets. Since only one pair
($\overline{h_1}$ and $h_{45}$) remains light and acquires
VEV, it turns out that families 1,2 and 3 get identified with the
$\tau$, $\mu$ and $e$-families respectively \cite{21}. The mass-heirarchy
and CKM mixings arise through higher dimensional
operators, by utilizing VEVs of appropriate fields and
hidden-sector condensates.

Including contributions from the entire massless spectrum,
one obtains: $Tr U_1 = Tr U_2 = Tr U_3 = 24$ and $Tr U_4 =
Tr U_5 = Tr U_6 = -12$. Thus, all six $U(1)_i$'s are
anomalous. They give rise to five anomaly-free combinations
and one anomalous one:
\bea
U'_1 &=& U_1 - U_2 \,,~~ \, U'_2 = U_4 - U_5 \,,~~ \,
 U'_3 = U_4 + U_5 - 2U_6 \,,\nn \\
\hat{U}_{\psi} &=& U_1 + U_2 - 2U_3 \,,\nn \\ 
\hat{U}_{\chi} &=& (U_1+U_2+U_3) + 2(U_4+U_5+U_6) \,,\nn \\
U_A &=& 2(U_1+U_2+U_3) - (U_4+U_5+U_6).
\tea
One obtains $Tr Q_A = 180$ \cite{21}. The anomalous $U_A$ is broken by
the Dine-Seiberg-Witten (DSW) mechanism \cite{26}, in which the
anomalous D-term generated by the VEV of the dilaton field
is cancelled by the VEVs of some massless fields which
break $U_A$, so that supersymmetry is preserved. The solutions
(i.e. the choice of fields with non-vanishing VEVs) to
the corresponding F and D - flat conditions are, however, not
unique. A few alternative possibilities have been
considered in Ref. \cite{21} (see also Refs. \cite{17} and \cite{24}
for analogous considerations).
Following our discussions in Sec. 2 as regards non-availibility
of $126$ of $SO(10)$ or ($1,3,\overline{10}$) of ${\cal G}_{224}$, we
assume, for the sake of simplicity in estimating strengths
of relevant operators,
that $B-L$ is violated
spontaneously at a scale $\sim 10^{15}$-$10^{16}$
 GeV by one unit (rather than two)
through the VEVs of {\it elementary} $s$neutrino-like
 fields $\widetilde{\overline{N_R}}
 \subset 16_H$ and $\widetilde{N_L'} \subset \overline{16_H}$
 (as in Ref. \cite{17}).
Replacing VEVs of these elementary
 fields by those of products of fields including condensates, 
as in Ref. \cite{21},
 would only lead to further suppression of the relevant unsafe
higher dimensional operators and go towards strengthening our
argument as
 regards certain symmetries being sufficient in preventing
rapid proton-decay \cite{27}.

%It is in order to remark here, however, that while the vector-like
%$(16_H + \overline{16_H})$ - pair, or equivalently the
%$(B=(1,2,\overline{4}) + \overline{B}'=(1,2,4))$ - pair, does not
%appear in the specific solution of Ref. \cite{20}, existence of such
%pairs is fairly generic in string theories, as in fact illustrated
%by the solutions of Refs. \cite{17} and \cite{18}. Furthermore, in Ref. \cite{17}, it is
%in fact the $s$neutrino-like fields which acquire VEVs through the
%DSW-mechanism, as assumed above.\footnote{It thus does not (at least) seem
%implausible that such vector-like $(16+\overline{16})$ - pairs
%can arise by suitable choice of boundary conditions for a variant
%of the solution of Ref. \cite{20}, as well.}
\noindent {\bf Proton-Decay Revisited:}
We now reexamine the problem of proton-decay and neutrino-masses
by assuming that in addition to $I_{3R}$ and
$B-L$, or just $Y$, either 
$\hat{Q}_{\psi} \equiv Q_1 + Q_2 - 2 Q_3$,
or $\hat{Q}_{\chi} \equiv Q_1+ Q_2 + Q_3 + 2( Q_4 + Q_5 + Q_6 ) $
(see Eq. (5)), 
or both
emerge as good symmetries near the string scale and that suitable
combinations of these symmetries (analogous to $Q_T = Q_{\psi}
-(B-L)$ of Sec. 2) survive up to some scale
$M_l \ll M_{st}$, even when $s$neutrino-like fields acquire VEVs $\sim 10^{15}$
- $10^{16}$ GeV. $M_l$ is determined in part by the VEVs of
electroweak doublets and singlets (denoted by $\phi$'s).
Generated radiatively, these are expected to be of order 1 TeV.
$M_l$ can also receive contributions from 
the hidden-sector condensates which can be much larger than 1 TeV.
As explained below, to ensure proton-stability, we need to assume that the
condensate-scale is $\leq 10^{-2.5} M_{st}$.
With the gauge coupling $\alpha_X$, at the unification-scale
$M_X$, having nearly the MSSM value of $.04 - .06$, or even an intermediate
value $\approx .16 - .2$ (say), as suggested in Ref. \cite{28},
this seems to be a safe assumption for most string models
(see discussions later).
The roles of the symmetries $Y$, $B-L$, $\hat{Q}_{\psi}$, $\hat{Q}_{\chi}$
and ($\hat{Q}_{\chi} + \hat{Q}_{\psi}$)
in allowing or forbidding the relevant $(B,L)$ - violating
operators, including the higher dimensional ones, which allow violations
of these symmetries through appropriate VEVs,
are shown in Table 3.
Based on the entries in this table, the following points are worth noting:

\noindent {\it (i)} \underline{Individual Roles of $\hat{Q}_{\psi}$
and $\hat{Q}_\chi$}: Once $Q_{\psi}$
of Sec. 2 is replaced by the anomaly-free but family-dependent
charge $\hat{Q}_{\psi} = Q_1 + Q_2 -2Q_3$, it no longer
forbids the $d=4$ operators $\Ubar \,\Dbar \,\Dbar$, $QL\Dbar$
and $LL\overline{E}$, when the three fields belong to three
{\it different} families. $\hat{Q}_{\psi}$ still forbids
the $d=5$ operators -- i.e. $QQQL/M$ and $\Ubar \,\Ubar \,\Dbar \,
\overline{E}/M$ etc. -- for all family - combinations.
 The charge $\hat{Q}_{\chi}$, which
is family - universal, forbids all three $d=4$ operators, but it
effectively allows them utilizing VEVs of $s$neutrino-like fields
through operators like
$\Ubar \,\Dbar \,\Dbar \langle\widetilde{\overline{N_R}}\rangle/M$
and $QL\Dbar \langle\widetilde{\overline{N_R}}\rangle/M$, which are unsafe. Furthermore, it allows
the $d=5$ operator $QQQL/M$, which is also unsafe.
We see from these
discussions and Table 3 that {\it no single charge provides
the desired protection against all the unsafe operators}. Let us
next consider pairs of charges.

\noindent {\it (ii)} \underline{Inadequacy of the Pairs ($Y$, $B-L$);
($Y$, $\hat{Q}_{\psi}$); ($Y$, $\hat{Q}_{\chi}$) and ($B-L$,
$\hat{Q}_{\chi}$)}: Table 3 \\
shows that the pair ($Y, B-L$)  does not give
adequate protection against some of the unsafe operators. Neither
do the pairs ($Y$, $\hat{Q}_{\psi}$), ($Y$, $\hat{Q}_\chi$) and
($B-L$, $\hat{Q}_\chi$). Amusingly enough, the charge
($\hat{Q}_{\chi}+ \hat{Q}_{\psi}$) by itself
forbids all unsafe operators except
when all four fields of the $d=5$ operator
$\Ubar \,\Ubar \,\Dbar \,\overline{E}/M$ belong to family 3. This
is unsafe if the identification of family 3 with the electron -
family \cite{21} is rigid.
We see that just one additional charge beyond $Y$ is not
adequate. Let us next consider other pairs of charges.

\noindent {\it (iii)} \underline{Adequate Protection Through the
Pair ($B-L$ and $\hat{Q}_{\psi}$) or the Pair
 ($\hat{Q}_{\chi}$ and $\hat{Q}_{\psi}$)}: \\
Using Table 3, we observe that
the pair ($B-L$ and $\hat{Q}_{\psi}$), as well as
the pair ($\hat{Q}_{\chi}$ and $\hat{Q}_{\psi}$), forbid
all unsafe operators, including
those which may arise from higher dimensional ones, with or without
hidden-sector condensates. In fact,
{\it members of the pairs mentioned above complement each other
in the sense that when one member of a pair allows an unsafe
operator, the other member of the same pair forbids it, and
vice versa} -- a remarkable team effort. 
Note that the strengths
of the d=4 and d=5 operators are controlled by the VEVs 
$ \, < \overline{h}_1/M >^2$, 
$< \Phi/M >^n$ and 
$< T_i \overline{T}_j/M^2 >^2$,
which give more than necessary suppression (see estimates below). 

\noindent {\it (iv)} \underline{$\hat{Q}_{\psi}$ removes Potential
Danger From Triplets in The Heavy Tower As Well}:
Color \\
triplets in the heavy infinite tower of
states with masses $M\sim M_{st} \sim 10^{18}$ GeV in general
pose a {\it potential danger} for all string theories, including
those for which they are projected out
from the massless sector \cite{21}. The exchange of these heavy
triplets, if allowed, would induce $d=5$ proton-decay operators
with strengths $\sim \kappa/M$, where $\kappa$ is given by
the product of two Yukawa couplings. Unless the Yukawa couplings are
appropriately suppressed \cite{29} so as to yield $\kappa \leq 10^{-7}$
\cite{3}, these operators would be unsafe. {\it Note, however,
that string-derived solutions possessing
symmetries like $\hat{Q}_{\psi}$ are free from this type
of danger}. This is because, if $\hat{Q}_{\psi}$ emerges as a good
symmetry near the string-scale, then the spectrum, the masses and the
interactions of the color-triplets in the heavy tower would respect
$\hat{Q}_{\psi}$. As a result, the exchange of such states can not
induce $d=5$ proton-decay operators, which violate $\hat{Q}_{\psi}$
(see Table 3).

In fact, for such solutions, the color-triplets in
the heavy tower can appear only as {\it vector-like pairs},
with opposite $\hat{Q}_{\psi}$-charges (like those in
$10$ and $\overline{10}$ of $SO(10)$, belonging to
$27$ and $\overline{27}$ of $E_6$ respectively), so that they can
acquire invariant masses of the type $M\{(H_3 \overline{H_3} +
H'_{3^\ast}\overline{H'_{3^\ast}}) + hc\}$, which conserve
$\hat{Q}_{\psi}$. Such mass-terms cannot induce
proton decay. By contrast, if only $I_{3R}$ and $B-L$, but not
$\hat{Q}_{\psi}$ (or something equivalent), emerged as good
symmetries, the mass-term
of the type $M(H_3 H'_{3^\ast} + hc)$ for the triplets in the heavy
tower would be permitted, which violates $\hat{Q}_{\psi}$ and can
in general induce proton-decay at an unacceptable rate.

Thus we see that a symmetry like $\hat{Q}_{\psi}$
plays an essential role in safegaurding proton-stability
from all angles.
%including contributions from the heavy tower of states. 
Since $\hat{Q}_{\psi}$
distinguishes between the three families \cite{30a}, it cannot, however,
arise within single - family grand unification
symmetries, including $E_6$. 
{\it But it does arise within string-derived
three-generation solutions (as in Ref. \cite{21}), which
at once know the existence of all three families}. In this
sense, string theory plays a vital role in explaining
naturally why the proton is so extraordinarily stable, in spite
of supersymmetry, and why the neutrinos are so light.

\section{$Z'$-mass and proton decay rate}
\baselineskip=20pt
If symmetries like $\hat{Q}_{\psi}$ and possibly
$\hat{Q}_\chi$, in addition to $I_{3R}$ and $B-L$, emerge as
good symmetries near the string scale, and break
spontaneously so that only electric
charge is conserved, there must exist at least one
extra $Z'$-boson (possibly more),
in addition to a superheavy $Z_H$ (that acquires mass when
$s$neutrino acquires a VEV) and the (almost) standard
$Z$ \cite{30}. The  extra $Z'$ boson(s) will be associated with
symmetries like $\hat{Q}_T \equiv 2\hat{Q}_{\psi} - (B-L)$ and
$\hat{Q}_\chi + \hat{Q}_{\psi}$, in addition to $Y$, that
survive after $s$neutrinos acquire VEVs. The $Z'$ bosons can
acquire masses through the VEVs of electroweak doublets and singlets
($\phi$'s), as well as
through the hidden-sector condensates like
$\langle\overline{T}_i T_j\rangle$, all of which break $\hat{Q}_T$
and $\hat{Q}_\chi + \hat{Q}_{\psi}$ (see Table 2).
As mentioned before, we
expect the singlet $\phi$'s to acquire VEVs, at least radiatively
(like the electroweak doublets), by utilizing their Yukawa couplings
with the doublets, which at the string-scale is comparable to
the top-Yukawa coupling (see Eq. (4)). Since the $\phi$'s do not
have electroweak gauge couplings, however, we would expect that
their radiatively-generated VEV, collectively denoted by $v_0$, to
be somewhat higher than those of the doublets ($v_{EW}\sim 200$ GeV)
-{\it i.e.}, quite
plausibly, $v_0 \sim 1$ TeV.
Ignoring possible contribution from the hidden sector,
we would thus expect the extra $Z'$ to be light
$\sim 1$ TeV.

To be specific, consider the case when
the string-scale symmetry (suppressing $SU(3)^C$ and $G_H$) is
given by ${\cal G}_{st} = {\cal G}_1 = SU(2)_L \times I_{3R} \times
(B-L) \times \hat{Q}_{\psi}$, which
breaks at a superheavy scale into $SU(2)_L \times Y \times
\left[\widetilde{Q_T}
= \hat{Q}_T + Y = 2\hat{Q}_{\psi} + (I_{3R} - (B-L)/2)\right]$
due to VEVs of $s$neutrino-like fields $\langle
\widetilde{\overline{N_R^i}}\rangle \neq 0$ (choose $i=1$ or
2, to be concrete). Now the VEVs of
electroweak doublets and singlets ($\leq {\cal O}(1$ TeV)) would
 break $SU(2)_L \times Y \times \tilde{Q}_T$
 to just $U(1)_{em}$. This second stage of SSB would
produce two relatively light $Z$-bosons, to be called $Z_1$ and $Z_2$.
$Z_1$ is the almost standard $Z$ with a mass
$\, = m_Z [1-{\cal O}(v_{EW}/v_0)^2]$;  $Z_2$ is the non-standard
$Z$ with a mass $M_{Z_2} =
(gv_0)[1+{\cal O}(v_{EW}/v_0)]$; the $Z$-$Z'$ mixing angle
is $\q \sim (v_{EW}/v_0)^2$.
Such a light $Z_2$ is compatible with known data \cite{31}, if
$v_0 \geq 10v_{EW} \approx 2$ TeV (say).
Alternative cases of ${\cal G}_{st}$
-- e.g. ${\cal G}_{st} = {\cal G}_2 = SU(2)_L \times Y \times
 \hat{Q}_{\psi} \times \hat{Q}_\chi$ and ${\cal G}_{st} = {\cal G}_3 =
SU(2)_L \times I_{3R} \times (B-L) \times \hat{Q}_{\psi} \times
\hat{Q}_\chi$ -- can be treated similarly. These will break
in the first step of SSB respectively into $SU(2)_L \times
Y \times (\hat{Q}_{\psi} + \hat{Q}_\chi)$ and $SU(2)_L \times [3$
orthogonal combinations of $Y, \hat{Q}_T$ and
$(\hat{Q}_{\psi} + \hat{Q}_\chi)]$, which in the second step
will produce one and two extra $Z'$-bosons, in
addition to
the almost standard $Z$. Details of this analysis
will be presented in a separate note.

If the hidden sector condensates like $\langle \overline{T}_i
T_j \rangle$ form, they would also contribute to the $Z'$-boson
masses. If the strength of $\langle \overline{T}_i
T_j \rangle$
is denoted by $\Lambda_c^2$, and if $\Lambda_c \sim \Lambda_H$,
where $\Lambda_H$ is the confinement-scale of the hidden-sector, their
contribution to $Z'$-mass, would typically far supercede that of the singlets,
because $\Lambda_H$ is expected to be superheavy
$\sim 10^{15}$-$10^{16}$ GeV, or at least medium-heavy $\sim 10^{8}$-$10^{13}$
GeV (see below). Nevertheless, with our present ignorance of the
hidden sector, 
it seems prudent to keep open
the possibility that its contribution to $Z'$-mass is
even zero \cite{32}, and that $Z'$ is light $\sim$ 1 TeV.

The mass of the $Z'$-boson is correlated with the
proton decay-rate. The heavier the $Z'$, the faster is the
proton-decay. Looking at Table 3, and allowing for the hidden
sector - condensates of strength $\Lambda_c^2$, we see that the strength
of the effective $d=4$ operators ($\Ubar \,\Dbar \,\Dbar $ etc.) is
given by
 $\left(\langle
\widetilde{\overline{N_R}}/M\rangle \right)\left(\langle T_i\overline{T}_j
\rangle /M^2\right)^2 \sim 10^{-2.5}(\Lambda_c/M)^4$, and
that of the $d=5$ operator ($QQQL/M$) is given by
$\left(\langle T_i\overline{T}_j
\rangle /M^2\right)^2 \sim (\Lambda_c/M)^4$.
The observed bound on the former ($\eta_{1,2} \leq 10^{-12}$)
implies a rough upper limit of $(\Lambda_c/M)^4\leq 10^{-9.5}$
and thus $\Lambda_c \leq 10^{15.5}$ GeV, while that on the
latter (i.e. $\lambda_{1,2} \leq 10^{-7}$) implies that
$\Lambda_c \leq 10^{16.2}$ GeV, where, for concreteness,
we have set $M = 10^{18}$
GeV.

Thus, if $\Lambda_c  \leq 1$ TeV \cite{32},  
$Z'$ would be light $\sim$ 
1 TeV, and accessible to 
LHC and perhaps NLC.
But, for this case, and even for $\Lambda_c \leq 10^{15}$ GeV (say),
proton-decay would be too slow ($\tau_p \geq 10^{42}$ yrs.) to be observed.
On the other hand, if
$\Lambda_c \sim 10^{15.4} - 10^{15.6}$ GeV,
the $Z'$-bosons would be
inaccessible; but proton decay would be observable
with a lifetime
$\sim 10^{32}$-$10^{35}$ years \cite{new33}. 
To see if such a superheavy $\Lambda_c$ is feasible, we note the
following. It has recently been suggested \cite{28}, that an intermediate
unified coupling $\alpha_X \approx 0.2$-$0.25$ at $M_X \sim
10^{17}$ GeV (as opposed to the MSSM-value of $\alpha_X \approx 1/26$)
is desirable to stabilize the dilaton and that such a value
of $\alpha_X$ would be realized if there exists a vector-like
pair of families having the quantum numbers of $16+\overline{16}$
of $SO(10)$, in the TeV-region. With
$\alpha_X \approx 0.16$-$0.2$ (say), and a hidden sector gauge symmetry
like $SU(4)_H$ or $SU(5)_H$ \cite{21}, 
a confinement scale
$\Lambda_H \sim \Lambda_c \sim 10^{15}-10^{16}$ GeV would in fact 
be expected. 
Thus, while
rapid proton decay is prevented by string-derived symmetries of the
type discussed here,
{\it observable rate for proton decay ($\tau_p \sim 10^{32}$-$10^{34}$
yrs.), which would be accessible to Superkamiokande and ICARUS,
seems perfectly natural
and perhaps called for} \cite{new33,new34}.

Before concluding, the following points are worth noting:

%\noindent \underline{(i) Effective $R$-Parity and LSP}: While
%$B-L$, $\hat{Q}_{\psi}$, $\hat{Q}_\chi$ and $\hat{Q}_T$ are broken
%through the VEVs of several fields and condensates -- i.e.
%$\widetilde{\overline{N_R^i}}$, $\langle T_i \overline{T_j} \rangle$,
%$\langle V_i \overline{V_j} \rangle$, $\overline{h_i}$,
%$h_{45}$ and the singlet $\phi$'s -- one can still define an
%effective $R$-parity given by $\hat{R} = (-1)^{4\hat{Q}_{\psi}}$.
%Noting that $\mid\hat{Q}_{\psi}\mid$ for all the fields and
%the condensates listed above, as well as for the quarks and the
%leptons of the three families, is given by $\frac{n}{2}$, where
%$n$ is an integer (see Table 2), it follows that $\hat{R}$ is
%$+1$ for all of them.
%Such an $R$-parity is thus conserved at all stages of SSB.
%It can not, of course, prevent the $d=4$ operators in Eq. (1),
%but, as discussed in Sec. 3, that task has been accomplished
%more efficiently by the pair symmetries ($B-L$ and $\hat{Q}_{\psi}$)
%or ($\hat{Q}_\chi$ and $\hat{Q}_{\psi}$). Defining $\tilde{R} =
%(-1)^{4\hat{Q}_{\psi}+2S}$, where $S$ denotes spin, which is also
%conserved, it thus follows that the lightest supersymmetric
%particle (LSP), which might be a neutralino, is stable,
%barring of course additional considerations pertaining to
%the mechanism for supersymmetry breaking which we have not
%dealt with. Such a stable LSP would be a good candidate for
%cold dark matter.

\noindent \underline{(i) The Messenger Sector}: The existence
of a messenger sector $G_M$, which is $[U(1)]^6$ for the
case considered here, is a generic property of a large class of
string-solutions (see e.g. \cite{21,17,18,24}). We have utilized this
sector to find symmetries like $\hat{Q}_{\psi}$ and $\hat{Q}_\chi$ to
prevent rapid proton decay. It is tempting to ask if the gauge
interactions of this sector, as opposed to standard model gauge
interactions \cite{33}, can help transmit SUSY-breaking efficiently
from the hidden to the observable sector and thereby ensure
$s$quark-degeneracy of at least the electron and the muon
families. This question will
be considered separately.

\noindent \underline{(ii) $\hat{Q}_{\psi}$ -- The Prototype of
A Desirable Symmetry}: $\hat{Q}_{\psi}$ is
a good example of the type of symmetry that can
safegaurd, in conjunction with $B-L$ or $\hat{Q}_\chi$, proton-stability
from
{\it all angles}, while permitting
neutrinos to have desired masses. It even helps eliminate the
potential danger from
contribution of the color-triplets in the heavy tower
of states.  In this
sense, $\hat{Q}_{\psi}$ plays a very desirable role. We do not,
however, expect it to be the
only choice. Rather, we expect other string-solutions to exist,
which would yield symmetries like $\hat{Q}_{\psi}$, serving
the same purpose \cite{34}. At the same time, we feel that {\it
emergence of symmetries like $\hat{Q}_{\psi}$ is a very
desirable constraint that should be built into the searches for
realistic string-solutions.}

To conclude, the following remark is in order. For the sake
of argument, one might have considered an $SO(10)$-type
SUSY grand unification by including $126$ of Higgs to break
$B-L$ and ignoring string-theory constraints \cite{16}. One
would thereby be able to forbid the $d=4$ operators and give desired masses
to the neutrinos \cite{15}. 
But, as mentioned before, the problems of finding a compelling
solution to the doublet-triplet splitting as well as to the
gravity-linked $d=5$ operators would still remain. This is
true not
just for SUSY $SO(10)$, but also for SUSY $E_6$, as well
as for the recently proposed SUSY $SU(5) \times SU(5)$ -
models \cite{35}. 
By contrast, a string-derived non-GUT model, possessing a symmetry like
$\hat{Q}_{\psi}$, in conjunction with $B-L$ or $\hat{Q}_\chi$, meets
naturally all the constraints
discussed in this letter. This shows that string theory
is not only needed for unity of all forces, but also for
ensuring {\it natural consistency} of SUSY-unification with two
low-energy observations --
proton stability and light masses for the neutrinos.\\

\noindent {\bf Acknowledgements} \\

I wish to thank M. Dine, L. Ibanez, G. Lazarides,
G. Leontaris, R.N. Mohapatra, A. Rasin, Q. Shafi, F. Wilczek, E. Witten 
and especially K.S. Babu, K. Dienes and A. Faraggi 
for most helpful discussions.

\begin{table}
\centering
\begin{tabular}{|c|c|c|c|c||c|c|c||c|c|}
\hline
Family & States & $Q_1$ & $Q_2$ & $Q_3$ & $Q_4$ & $Q_5$ & $Q_6$ &
$\hat{Q}_{\psi}$ & $\hat{Q}_{\chi}$ \\ \hline
 & $q_1$ & $1/2$ & $0$ & $0$ & -$1/2$ & $0$ & $0$ & $1/2$
 & -$1/2$ \\
$1$ & $L_1$ & $1/2$ & $0$ & $0$ & $1/2$ & $0$ & $0$ & $1/2$
& $3/2$ \\
 & $(\Ubar , \overline{E})_1$ & $1/2$ & $0$ & $0$ & $1/2$ & $0$ & $0$ &
 $1/2$ & $3/2$ \\
  & $(\Dbar , \overline{\nu_R})_1$ & $1/2$ & $0$ & $0$ & -$1/2$ & $0$ & $0$ &
 $1/2$ & -$1/2$ \\  \hline
  & $q_2$ & $0$ & $1/2$ & $0$ & $0$ & -$1/2$ & $0$ & $1/2$
  & -$1/2$ \\
$2$ & $L_2$ & $0$ & $1/2$ & $0$ & $0$ & $1/2$ & $0$ & $1/2$
  & $3/2$ \\
 & $(\Ubar , \overline{E})_2$ & $0$ & $1/2$ & $0$ & $0$ & $1/2$ & $0$ & $1/2$
  & $3/2$ \\
 & $(\Dbar , \overline{\nu_R})_2$ & $0$ & $1/2$ & $0$ & $0$ & -$1/2$ & $0$ & $1/2$
  & -$1/2$ \\ \hline
 & $q_3$ & $0$ & $0$ & $1/2$ & $0$ & $0$ & -$1/2$ & -$1$ &
 -$1/2$ \\
$3$ & $L_3$ & $0$ & $0$ & $1/2$ & $0$ & $0$ & $1/2$ & -$1$ &
 $3/2$ \\
 & $(\Ubar , \overline{E})_3$ & $0$ & $0$ & $1/2$ & $0$ & $0$ & $1/2$ & -$1$ &
 $3/2$ \\
 & $(\Dbar , \overline{\nu_R})_3$ & $0$ & $0$ & $1/2$ & $0$ & $0$ & -$1/2$ & -$1$ &
 -$1/2$ \\   \hline \hline
 & & & & & & & & & \\
Color & $D_{45} = ({\bf 3},-2/3, 1_L, 0)$ & -$1/2$ &
-$1/2$ & $0$ & $0$ & $0$ & $0$ & -$1$ & -$1$ \\
Triplets & $\overline{D_{45}} = ( 3^{\ast},+2/3, 1_L, 0)$ & $1/2$ &
$1/2$ & $0$ & $0$ & $0$ & $0$ & $+1$ & $+1$ \\
 & & & & & & & & & \\ \hline
 & & & & & & & & & \\
 & $\overline{h_1} = ({\bf 1},0,{\bf 2_L},1/2)$ & -$1$ &
$0$ & $0$ & $0$ & $0$ & $0$ & -$1$ & -$1$ \\
Higgs & $\overline{h_2} = ({\bf 1},0,{\bf 2_L},1/2)$ & $0$ &
-$1$ & $0$ & $0$ & $0$ & $0$ & -$1$ & -$1$ \\
doublets & $\overline{h_3} = ({\bf 1},0,{\bf 2_L},1/2)$ & $0$ &
$0$ & -$1$ & $0$ & $0$ & $0$ & $+2$ & -$1$ \\
 & $\overline{h_{45}} = ({\bf 1},0,{\bf 2_L},1/2)$ & $1/2$ &
$1/2$ & $0$ & $0$ & $0$ & $0$ & $1$ & $1$ \\
 & & & & & & & & & \\ \hline
 & & & & & & & & & \\
 & $V_1, \overline{V}_1$ & $0$ & $1/2$ & $1/2$ & $1/2$ & $0$ & $0$ &
 -$1/2$ & $2$ \\
 & $T_1, \overline{T}_1$ & $0$ & $1/2$ & $1/2$ & -$1/2$ & $0$ & $0$ &
 -$1/2$ & $0$ \\ \cline{2-10}
 & & & & & & & & & \\
Hidden & $V_2, \overline{V}_2$ & $1/2$ & $0$ & $1/2$ & $0$ & $1/2$ & $0$ &
 -$1/2$ & $2$ \\
Matter & $T_2, \overline{T}_2$ & $1/2$ & $0$ & $1/2$ & $0$ & -$1/2$ & $0$ &
 -$1/2$ & $0$ \\ \cline{2-10}
 & & & & & & & & & \\
 & $V_3, \overline{V}_3$ & $1/2$ & $1/2$ & $0$ & $0$ & $0$ & $1/2$ &
 $1$ & $2$ \\
 & $T_3, \overline{T}_3$ & $1/2$ & $1/2$ & $0$ & $0$ & $0$ & -$1/2$ &
 $1$ & $0$ \\ \hline
\end{tabular}
\caption{Partial List of Massless States from Ref. [20].
(i) The quark and lepton fields have the standard properties
under $SU(3)^C \times U(1)_{B-L} \times SU(2)_L \times U(1)_{I_{3R}}$,
which are not shown, but those of color triplets and Higgses are
shown. (ii) Here $\hat{Q}_{\psi} \equiv Q_1 + Q_2 -2Q_3$
and $\hat{Q}_{\chi} = (Q_1 + Q_2 + Q_3) + 2(Q_4 + Q_5 + Q_6)$ (see Eq. (5)).
(iii) The doublets $\overline{h}_{1,2,3,45}$ are
accompanied by four doublets $h_{1,2,3,45}$ with quantum numbers
of conjugate representations, which are not shown. (iv) The
$SO(10)$-singlets 
$ \{ \phi \} $ 
which possess $U(1)_i$-charges,
 and the fractionally charged states which
become superheavy, or get confined [24], are not shown.
In Ref. [20], since only $\overline{h_1}$ and $h_{45}$ remain
light,  families
1, 2 and 3
get identified with the $\tau$, $\mu$ and $e$ - families respectively.
Hidden matter $V_i, \overline{V}_i, T_i$ and $\overline{T}_i$
are $SO(10)$-singlets
and transform as $(1,{\bf 3}), (1, \overline{{\bf 3}}), ({\bf 5}, 1)$ and
$(\overline{{\bf 5}}, 1)$, respectively, under $SU(5)_H \times SU(3)_H$.}
\end{table}
\begin{table}
\centering
\begin{tabular}{|c|c|c|c|c|c|c|c|}
\hline
Operators & Family  & $Y$ & $B-L$ & $\hat{Q}_{\psi}$ &
$\hat{Q}_{\chi}$ & $\hat{Q}_{\chi}+ \hat{Q}_{\psi}$ & If  \\
 & Combinations & & & & & & Allowed \\ \hline
 & (a) All & & & & & & \\
$\Ubar \,\Dbar \,\Dbar , QL\Dbar , LL\overline{E}$
  & except (b)
& $\surd$ & $\times$ & $\times$ & $\times$ & $\times$ &
unsafe \\ \cline{2-8}
  & (b) 3 fields from  & & & & & & \\
 & 3 different & $\surd$ & $\times$ &
  $\surd$ & $\times$ & $\times$ &
unsafe  \\
 & families & & & & & & \\ \hline
 & & & & & & & \\
$(\Ubar \,\Dbar \,\Dbar$ or $QL\Dbar)(\overline{N_R}/M)$
& All & $\surd$ & $\surd$ & $\times$ & $\surd$ & $\times$
& unsafe \\
 & & & & & & & \\
$(\Ubar \,\Dbar \,\Dbar$ or $QL\Dbar) (\overline{N_R}/M)\times$
& All & $\surd$ & $\surd$
& $\surd$ &  $\times$ & $\times$ & safe \\
$[(\overline{h_1}/M)^2$ or (``$\phi$''$/M)^n]$ & & & & & & & \\
 & & & & & & & \\
$(\Ubar \,\Dbar \,\Dbar$ or $QL\Dbar) (\overline{N_R}/M)\times$
& Some($\dagger$) & $\surd$ & $\surd$ & $\surd$ & $\surd$ & $\surd$
& safe \\
$(T_i \overline{T}_j/M^2)^2$ & & & & & & & \\ \hline
 & & & & & & & \\
$QQQL/M$ & All & $\surd$ & $\surd$ & $\times$ & $\surd$ & $\times$
& unsafe  \\
$(QQQL/M)(N_L^i/M)_{i=1,2}$ & e.g.($1,2,1,3$) & $\surd$ & $\times$
& $\surd$ & $\times$ & $\times$ & unsafe  \\
$(QQQL/M)(N_L^i/M)(\overline{N}_R^j/M)$ & All & $\surd$ & $\surd$
& $\times$ & $\surd$ & $\times$ & safe(?)  \\
 & & & & & & & \\
$(QQQL/M)(T_i\overline{T}_j/M^2)^2$ & Some($\dagger$) & $\surd$ & $\surd$
& $\surd$ & $\surd$ & $\surd$ & safe  \\
$\Ubar \,\Ubar \,\Dbar \,\overline{E}/M$ & All & $\surd$ & $\surd$
& $\times$ & $\times$ & $(\ast)$ & unsafe  \\
$LL\overline{h_i}\,\overline{h_i}/M$ & All & $\surd$ & $\times$
& $\times$ & & $(\ast)$ & safe  \\
 & & & & & & & \\ \hline
\end{tabular}
\caption{The roles of $Y$, $B-L$, $\hat{Q}_{\psi}$,
$\hat{Q}_{\chi}$ and $\hat{Q}_{\chi}+ \hat{Q}_{\psi}$ in allowing or forbidding the
relevant $(B,L)$ violating operators. Check mark ($\surd$) means
``allowed'' and cross ($\times$) means ``forbidden''. The mark
$\dagger$ signifies that the corresponding operator is allowed if
either two of the four fields are in
family (1 or 2) and two are in family 3, with $i=1$ and $j=3$; or
all four fields are in family (1 or 2) with $i=1$ and $j=2$. The mark
$(\ast)$ signifies that $(\hat{Q}_{\chi}+ \hat{Q}_{\psi})$ forbids
$\Ubar \,\Ubar \,\Dbar \,\overline{E}/M$ for all family-combinations
except when all four fields belong to family 3, and that it forbids
$LL\overline{h_i}\,\overline{h_i}$ in some family-combinations,
but not in others. In labelling the operators as safe/unsafe, we
have assumed that $\langle\widetilde{\overline{N_R^i}}\rangle\sim 10^{15.5}$
GeV, $\langle\phi/M\rangle^n \leq 10^{-9}$ and $M\sim M_{st} \sim 10^{18}$ GeV,
and that hidden sector condensate-scale $\Lambda_c \leq
10^{15.5}$ GeV (see text). Note that the pairs ($Y$, $B-L$),
$(Y$, $\hat{Q}_{\psi})$, $(Y$, $\hat{Q}_{\chi})$ and $(B-L$, $\hat{Q}_{\chi})$
do not give adequate
protection against the unsafe operators. But $\hat{Q}_{\psi}$, in
conjunction with $B-L$ or $\hat{Q}_\chi$, gives adequate protection against
all unsafe operators.
This establishes the necessity of string-derived symmetries like
$\hat{Q}_{\psi}$ (which can not emerge
from familiar GUTs including $E_6$) in ensuring proton-stability.}
\end{table}


\begin{thebibliography}{999}
\bibitem{29} As far as we can see, Yukawa couplings of the color-triplets
     in the heavy tower need not
     all be suppressed to the same extent as those of the
     electroweak doublets.
\bibitem{30a} Note $\hat{Q}_{\psi}$ assigns the same charge to all 
     sixteen members of a given family. In this sense,
$\hat{Q}_\psi$ (though not $\hat{Q}_\chi$) commutes with $SO(10)$.
\bibitem{32} This would be the case if, for  a certain
     string-solution, 
the hidden-sector condensates, even if they
     form, do not break the corresponding proton-stabilizing symmetry.
     In this case, they will not contribute to either the $Z'$-mass
     or to the strength of the proton-decay operator.
\bibitem{new33} If $< \widetilde{\overline{N}}_R > \neq 0$, 
the d=4 operators (see table 3)
lead to proton decay rate $\propto
( \eta_1 \eta_2 )^2 \propto (\Lambda_c / M )^{16}$, which varies 
extremely rapidly with
$\Lambda_c$. It is interesting that, despite this rapid variation,
proton decay is observable for a plausible 
though narrow range of values of $\Lambda_c \approx 10^{15.5}$ GeV. 
If $< \widetilde{\overline{N}}_R > = 0$, as in Ref. \cite{21} (see also
remarks in Ref. \cite{27}), only the d=5 operators with strengths
$\sim ( < T_i \overline{T}_j > / M^2 )^2 $ are relevant (see table 3).
These will lead to proton decay rate $\propto ( \Lambda_c / M )^8$,
which would be observable if $\Lambda_c \sim 10^{16}$ GeV (see text).
\bibitem{new34} In all these considerations, contribution of d=6
operators are neglected, because they are strongly suppressed if the relevant
scale ($M_X$ or $M$) exceeds $10^{16}$ GeV.
\end{thebibliography}
\end{document}